\documentclass[10pt,ps]{iopart}

\usepackage{graphicx}

\begin{document}

\title[]{Coherence and mode decomposition of weak twin beams}

\author{Jan Pe\v{r}ina Jr.}
\address{RCPTM, Joint Laboratory of Optics of Palack\'{y} University
and Institute of Physics of Academy of Sciences of the Czech
Republic, 17. listopadu 12, 771~46 Olomouc, Czech Republic}
\ead{jan.perina.jr@upol.cz}

\begin{indented}
\item[]
\end{indented}

\begin{abstract}
Properties of weak spatio-spectral twin beams in paraxial
approximation are analyzed using the decomposition into
appropriate paired modes. Numbers of paired modes as well as
numbers of modes in the signal (or idler) field in the transverse
wave-vector and spectral domains are analyzed as functions of
pump-beam parameters. Spatial and spectral coherence of weak twin
beams is described by auto- and cross-correlation functions.
Relation between the numbers of modes and coherence is discussed.
\end{abstract}


\vspace{2pc} \noindent{\it Keywords}: twin beam, spatial and
spectral coherence, number of modes

\maketitle

\ioptwocol

\section{Introduction}

Generation of photon pairs by spontaneous parametric
down-conversion \cite{Boyd2003} belongs to the most frequently
studied nonlinear processes in optics. The reason is entanglement
of photons constituting a photon pair. It represents a purely
quantum property. Entanglement leads to nonclassical
cross-correlation functions between the signal and idler fields
\cite{Mandel1995} that influence many physical effects.
Nonclassical multiple coincidence-count rates used, e.g., for
tests of the Bell inequalities \cite{Weihs1998,Genovese2005} or
quantum teleportation \cite{Bouwmeester1997} represent the most
important manifestation of these correlations. Unusual entangled
two-photon absorption with its entanglement-induced transparency
\cite{Fei1997,PerinaJr1998} and virtual state two-photon
spectroscopy \cite{Saleh1998} is another example of the large
impact of entanglement of photons. Also ghost imaging has to be
mentioned here \cite{Gatti2008}. At present, there already exist
several applications based upon entangled photon pairs including
quantum key distribution \cite{Gisin2002}, ultra-fast measurements
\cite{Carrasco2004} and absolute detector calibration
\cite{Klyshko1980,PerinaJr2012a}.

This wide use of entangled photon pairs has naturally stimulated
investigations of a detailed structure of photon pairs. It has
been shown that their properties are fully described by two-photon
(spectral, temporal, spatial) amplitudes at the level of
individual photon pairs \cite{Rubin1994,PerinaJr1999a}. Moreover,
the Schmidt decomposition of a two-photon amplitude
\cite{Law2000,Law2004} has been found crucial for the
quantification of entanglement by the Schmidt number. We note that
no physical quantity is directly related to entanglement but
entanglement influences even qualitatively the physical behavior
of photon pairs. As the experimental determination of the Schmidt
number \cite{Chan2007,Just2013} and profiles of the modes
\cite{Christ2011,Bobrov2013,Avella2011} is difficult, an
alternative approach based on the measurement of field width and
width of the corresponding intensity cross-correlation function
has been developed giving the Fedorov ratio
\cite{Fedorov2005,Mikhailova2008,Brida2009c} as a good quantifier
of entanglement.

In the last years, a greater deal of attention has been devoted to
fields composed of many photon pairs arising in parametric
down-conversion \cite{Haderka2005a,Bondani2007,Jedrkiewicz2012}.
Properties of these twin beams reflect those of the individual
photon pairs \cite{Caspani2010}, at least for not very intense
twin beams \cite{Allevi2014a}. For characterizing twin beams,
auto-correlation functions are important as well as the
cross-correlation functions \cite{Machulka2014}. Auto-correlation
functions then give, according to statistical optics
\cite{Perina1985}, the information about the number of modes
constituting the signal (or idler) part of the twin beam. In
analogy with the definition of Fedorov ratio, this number of modes
can be determined by the ratio of field width and width of the
corresponding auto-correlation function. Considering weak twin
beams, the auto-correlation functions can even be derived from the
two-photon amplitudes describing photon pairs.

In this contribution, we determine side by side the spatial and
spectral auto- and cross-correlation functions as they depend on
pump-field parameters. Using these functions, we obtain the number
of modes in the signal (or idler) field and compare it with the
Schmidt number giving the number of paired modes
\cite{Mikhailova2008}.

The paper is organized as follows. Theory suitable for describing
weak twin beams is presented in Sec.~2. Sec.~3 is devoted to the
properties of twin beams in the transverse wave-vector plane.
Spectral properties of the twin beams are discussed in Sec.~4.
Sec.~5 brings conclusions.

\section{Theory of weak twin beams}

Parametric down-conversion in a medium with tensor $ d $ of the
second-order nonlinear coefficients  is characterized by momentum
operator $ \hat{G}_{\rm int} $ written as follows
\cite{Perina1991,PerinaJr2000}:
\begin{eqnarray}   
 \hat{G}_{\rm int}(z) &=& 4 \epsilon_0 \int dxdy \int_{-\infty}^{\infty} dt \nonumber \\
 & &  \hspace{-1mm} \left[
  {d} : E^{(+)}_p({\bf r},t) \hat{E}^{(-)}_s({\bf r},t)
  \hat{E}^{(-)}_i({\bf r},t) + {\rm h.c.} \right];
\label{1}
\end{eqnarray}
$ {\bf r} = (x,y,z) $. Symbol $ E^{(+)}_p $ denotes the
positive-frequency part of the classical pump electric-field
amplitude and $ \hat{E}^{(-)}_s $ [$ \hat{E}^{(-)}_i $] means the
negative-frequency part of the signal- [idler-] field operator
amplitude. Permittivity of vacuum is denoted as $ \epsilon_0 $,
symbol : is shorthand for tensor shortening with respect to its
three indices and $ {\rm h.c.} $ replaces the Hermitian conjugated
term.

Amplitudes of all three interacting fields can be decomposed into
harmonic plane waves with wave vectors $ {\bf k}_a $ and
frequencies $ \omega_a $:
\begin{eqnarray}   
 E^{(+)}_a({\bf r},t) &=& \frac{1}{\sqrt{2\pi}^3} \int
  d{\bf k}_a \, E^{(+)}_a({\bf k}_a) \exp(i{\bf k}_a{\bf r} -i\omega_a
  t), \nonumber \\
 & & \hspace{1cm} a=p,s,i.
\label{2}
\end{eqnarray}
In the considered paraxial approximation, the plane wave with wave
vector $ {\bf k}_a $ is conveniently parameterized by its
frequency $ \omega_a $ and transverse wave vector $ {\bf
k}^\perp_a $. The quantum signal and idler spectral electric-field
amplitudes $ \hat{E}_a^{(-)}({\bf k}_a^\perp,\omega_a) $ can then
be expressed in terms of the appropriate creation operators $
\hat{a}^\dagger({\bf k}_a^\perp,\omega_a) $:
\begin{equation}  
 \hat{E}_a^{(-)}({\bf k}_a^\perp,\omega_a) = -i \sqrt{
  \frac{\hbar\omega_a^2}{2\epsilon_0 c^2 k_a} } \,
  \hat{a}_a^{\dagger}({\bf k}_a^\perp,\omega_a) ;
\label{3}
\end{equation}
$ \hbar $ is the reduced Planck constant and $ c $ is the speed of
light in vacuum; $ k_a= |{\bf k}_a| $.

Correlations in the transverse wave-vector planes of the signal
and idler fields are described by the following function $ T_L $
depending on the pump-field transverse spatial spectrum $
E_p^\perp({\bf k}_p^\perp) $:
\begin{eqnarray}  
 T_L({\bf k}_s^\perp,{\bf k}_i^\perp) &=& E_p^\perp({\bf k}_s^\perp+{\bf k}_i^\perp) \nonumber \\
 & &  \hspace{-2.2cm} \mbox{} \times \exp \left( -i\left[ \frac{|{\bf k}_s^\perp+{\bf k}_i^\perp|^2}{2k_p}
  -\frac{|{\bf k}_s^\perp|^2}{2k_s} - \frac{|{\bf k}_i^\perp|^2}{2k_i}
  \right] \frac{L}{2} \right) \nonumber \\
 & &  \hspace{-2.2cm} \mbox{} \times {\rm sinc} \left( \left[ \frac{|{\bf k}_s^\perp+{\bf k}_i^\perp|^2}{2k_p}
  -\frac{|{\bf k}_s^\perp|^2}{2k_s} - \frac{|{\bf k}_i^\perp|^2}{2k_i}
  \right] \frac{L}{2} \right);
\label{4}
\end{eqnarray}
$ {\rm sinc}(x) \equiv \sin(x)/x $ and $ \delta $ stands for the
Dirac $ \delta $-function. In Eq.~(\ref{4}), $ |{\bf k}_a^\perp|^2
= k_{a,x}^2 + k_{a,y}^2 $ and $ L $ denotes the crystal length.

We assume that the emitted signal and idler fields have the radial
symmetry. This is a good approximation for sufficiently narrow
spatial spectral profiles $ E_p^\perp $ of the pump field. The
pump field in the considered type-I nonlinear interaction
propagates as an extraordinary wave and so its index $ n_p $ of
refraction changes linearly with the radial emission angle $
\vartheta_p $ [$ \vartheta_p = \arcsin(|{\bf k}_p^\perp|/k_p) $]
in the plane containing the propagation direction and the crystal
optical axis. This dependence introduces anisotropy into the
generated signal and idler fields
\cite{Fedorov2008,Fedorov2007,Osorio2007}. Assuming the Gaussian
transverse pulse profile with radius $ w_p $ [$ E_p^\perp(|{\bf
k}_p^\perp|) = w_p/\sqrt{2\pi} \exp( - w_p^2|{\bf k}_p^\perp|^2/4)
$] the anisotropy is well developed for narrow transverse pump
profiles with wide spatial spectra. Anisotropy considerably
modifies the emitted signal and idler fields for the pump-field
radii $ w_p $ comparable or smaller than $ w_p^a $ (for more
details, see \cite{Fedorov2008}):
\begin{equation} 
 w_p^a = \frac{1}{n_p} \left|\frac{d n_p(\omega_p^0,\vartheta_p)}{d\vartheta_p}
  \right| \frac{L}{x_e} ;
\label{5}
\end{equation}
$ x_e \approx 2.2 $ [$ \sin(x_e)/x_e = 1/e $].

The assumed radial symmetry qualitatively simplifies the
description as it allows to decompose the function $ T_L $ into
the dual Schmidt basis in both radial and azimuthal directions in
the transverse planes:
\begin{eqnarray}  
 T_L({\bf k}_s^\perp,{\bf k}_i^\perp) &=&
  \frac{t^\perp}{2\pi \sqrt{k_s^\perp k_i^\perp} } \sum_{m=-\infty}^{\infty}
  \sum_{l=0}^{\infty} \lambda_{ml}^\perp  \nonumber \\
 & & \mbox{} \times t_{s,ml}(k_s^\perp,\varphi_s)
  t_{i,ml}(k_i^\perp,\varphi_i),
\label{6}  \\
 & & \hspace{-12mm} t_{s,ml}(k_s^\perp,\varphi_s) = u_{s,ml}(k_s^\perp) \exp( i m
  \varphi_s), \nonumber \\
 & & \hspace{-12mm} t_{i,ml}(k_i^\perp,\varphi_i) = u_{i,ml}(k_i^\perp) \exp( -i m
  \varphi_i). \nonumber
\end{eqnarray}
Whereas the functions $ u_{a,ml} $ describe the radial parts of
the fields, the harmonic functions $ \exp( i m \varphi_a) /
\sqrt{2\pi} $ are appropriate for the azimuthal parts of the
fields, $ a=s,i $. Symbols $ \lambda_{ml}^\perp $ denote the
Schmidt numbers and $ t^\perp $ is the normalization constant.

Unitary transformations of the field operators,
\begin{eqnarray}  
 \hat{a}_{a,ml}(\omega_a,z) &=& \int_{0}^{\infty} dk_a^\perp \int_{0}^{2\pi}
  d\varphi_a \, t_{a,ml}^*(k_a^\perp,\varphi_a) \nonumber \\
 & & \mbox{} \times  \hat{a}_{a}(k_a^\perp,\varphi_a,\omega_a,z),  \hspace{3mm} a=s,i,
\label{7}
\end{eqnarray}
then allow us to write the first-order perturbation solution of
the Schr\"{o}dinger equation as follows:
\begin{eqnarray}  
 |\psi\rangle_{\rm out} &=& t^\perp \sum_{m,l} \lambda_{ml}^\perp
  \int_{0}^{\infty} d\omega_s \int_{0}^{\infty} d\omega_i \, F_L(\omega_s,\omega_i)
  \nonumber \\
 & & \mbox{} \times
  \hat{a}_{s,ml}^{\dagger}(\omega_s,0)
   \hat{a}_{i,ml}^{\dagger}(\omega_i,0) |{\rm vac}\rangle;
\label{8}
\end{eqnarray}
$ |{\rm vac}\rangle $ is the incident vacuum state. The two-photon
spectral amplitude $ F_L $ introduced in Eq.~(\ref{8}) is
determined by the formula
\begin{eqnarray}  
 F_L(\omega_s,\omega_i) &=& \frac{2id_{\rm eff}L}{\sqrt{2\pi}^3 c^2}
  \frac{\omega_s\omega_i}{\sqrt{k_s k_i}} E_p^\parallel
  (\omega_s+\omega_i) \nonumber \\
 & & \hspace{-1.7cm} \mbox{} \times
  \exp \left(-i[k_p(\omega_s+\omega_i)-k_s(\omega_s)-k_i(\omega_i)]L/2\right)
  \nonumber \\
 & & \hspace{-1.7cm} \mbox{} \times
  {\rm sinc} \left(
  [k_p(\omega_s+\omega_i)-k_s(\omega_s)-k_i(\omega_i)]L/2\right),
\label{9}
\end{eqnarray}
where $ E_p^\parallel $ stands for the pump-field spectrum and $
d_{\rm eff} $ denotes an effective nonlinear coupling constant.
The Schmidt decomposition of two-photon amplitude $ F_L $,
\begin{equation}  
 F_L(\omega_s,\omega_i) = f^\parallel \sum_{q=0}^{\infty}
 \lambda_q^\parallel f_{s,q}(\omega_s) f_{i,q}(\omega_i),
\label{10}
\end{equation}
and introduction of new field operators $ \hat{a}_{a,mlq} $,
\begin{equation}  
 \hat{a}_{a,mlq} = \int_{0}^{\infty} d\omega_a \, f_{a,q}^*(\omega_a)
  \hat{a}_{a,ml}(\omega_a,0) ,  \hspace{3mm} a=s,i,
\label{11}
\end{equation}
allows to rearrange the output state $ |\psi\rangle_{\rm out} $
into the form:
\begin{eqnarray}  
 |\psi\rangle_{\rm out} = t^\perp f^\parallel \sum_{m,l,q} \lambda_{ml}^\perp
  \lambda_q^\parallel \hat{a}_{s,mlq}^{\dagger}
   \hat{a}_{i,mlq}^{\dagger} |{\rm vac}\rangle .
\label{12}
\end{eqnarray}
Formula (\ref{12}) represents the output state $ |\psi\rangle_{\rm
out} $ decomposed into independent paired spatial and spectral
modes. We note that there occurs degeneracy in the signal and
idler mode structure for $ m=0 $ owing to the radial and spectral
symmetry of the signal and idler fields. Due to this degeneracy,
we cannot distinguish a signal and an idler photons in modes with
$ m=0 $ as both photons are created in one spatio-spectral mode.
This considerably modifies properties of the generated paired
fields in the collinear geometry \cite{Fedorov2014}. On the other
hand, non-collinear geometries are practically unaffected by this
degeneracy as the relative contribution of such photon pairs to
the structure of a twin beam is negligible.

The emitted signal (or idler) field is characterized by its
intensity profiles along variables $ k_a^\perp $, $ \varphi_a $
and $ \omega_a $, $ a=s,i $. Its internal correlations are
described by amplitude correlation functions in these variables.
The mutual signal and idler correlations are characterized by
intensity cross-correlation functions in these variables. In
detail, intensity profile $ n_{s,k} $ (expressed in photon-number
density) [$ n_{s,k}(k_s^\perp) \approx \langle
\hat{a}_s^\dagger(k_s^\perp) \hat{a}_s(k_s^\perp) \rangle $] of
the signal field in the radial wave-vector direction is derived
from the function $ T_L $ given in Eq.~(\ref{6}):
\begin{eqnarray}  
 n_{s,k}(k_s^\perp) &=& k_s^\perp \int_{0}^{\infty} dk_i^\perp k_i^\perp
  \nonumber \\
 & & \mbox{} \times | T_{L}(k_s^\perp,\varphi_s=0,k_i^\perp,\varphi_i=\pi)|^2.
\label{13}
\end{eqnarray}
Using Eq.~(\ref{9}) the signal-field intensity spectrum $
n_{s,\omega} $ [$ n_{s,\omega}(\omega_s) \approx \langle
\hat{a}_s^\dagger(\omega_s) \hat{a}_s(\omega_s) \rangle $] is
given by the formula
\begin{equation}  
 n_{s,\omega}(\omega_s) = \int_{0}^{\infty} d\omega_i \,
  | F_{L}(\omega_s,\omega_i)|^2.
\label{14}
\end{equation}

The amplitude signal-field correlations in their radial [$
A_{s,k}^a $, $ A_{s,k}^a(k_s^\perp,k_s^{'\perp}) \approx \langle
\hat{a}_s^\dagger(k_s^\perp) \hat{a}_s(k_s^{'\perp}) \rangle $],
azimuthal [$ A_{s,\varphi}^a $] and spectral [$ A_{s,\omega}^a $]
variables are described by the corresponding auto-correlation
functions given as:
\begin{eqnarray}  
 A_{s,k}^a(k_s^\perp,k_s^{'\perp}) &=& \sqrt{k_s^\perp k_s^{'\perp}}
  \int_{0}^{\infty} dk_i^\perp \,
  k_i^\perp  T_{L}^*(k_s^\perp,\varphi_s^0=0,  \nonumber \\
 & & \hspace{-5mm} k_i^{\perp},\varphi_i^0=\pi) \,
  T_{L}(k_s^{'\perp},\varphi_s^0=0,
  k_i^{\perp},\varphi_i^0=\pi), \nonumber \\
 A_{s,\varphi}^a(\varphi_s,\varphi'_s) &=& k_s^{\perp 0}
  k_i^{\perp 0}
  \int_{0}^{2\pi} d\varphi_i \, T_{L}^*(k_s^{\perp 0},\varphi_s,
  k_i^{\perp 0},\varphi_i) \nonumber \\
 & & \mbox{} \times  T_{L}(k_s^{\perp 0},\varphi'_s,
  k_i^{\perp 0},\varphi_i), \nonumber \\
 A_{s,\omega}^a(\omega_s,\omega'_s) &=& \int_{0}^{\infty} d\omega_i \, F_{L}^*(\omega_s,\omega_i)
  F_{L}(\omega'_s,\omega_i).
\label{15}
\end{eqnarray}

The intensity correlations between the signal and idler fields in
their radial [$ C_{s,k} $, $ C_{s,k}(k_s^\perp,k_i^\perp) \approx
\langle {\cal N}:
\hat{a}_s^\dagger(k_s^\perp)\hat{a}_s(k_s^{\perp})
\hat{a}_i^\dagger(k_i^{\perp}) \hat{a}_i(k_i^{\perp}):\rangle $,
symbol $ {\cal N}:: $ mean the normal ordering of field
operators], azimuthal [$ C_{s,\varphi} $] and spectral [$
C_{s,\omega} $] variables are characterized by the following
cross-correlation functions:
\begin{eqnarray}  
 C_{s,k}(k_s^\perp,k_i^{\perp}) &=& k_s^\perp k_i^{\perp} | T_{L}(k_s^\perp,\varphi_s^0=0,
  k_i^{\perp},\varphi_i^0=\pi)|^2, \nonumber \\
 C_{s,\varphi}(\varphi_s,\varphi_i) &=& k_s^{\perp 0} k_i^{\perp 0} | T_{L}(k_s^{\perp 0},\varphi_s,
  k_i^{\perp 0},\varphi_i)|^2, \nonumber \\
 C_{s,\omega}(\omega_s,\omega_i) &=&  | F_{L}(\omega_s,\omega_i)|^2.
\label{16}
\end{eqnarray}
Similar quantities as written in Eqs.~(\ref{13}---\ref{16}) for
the signal field can be defined also for the idler field.

The number of effectively populated paired modes in a twin beam is
determined by the Schmidt number $ K $ defined as
\cite{URen2003,Gatti2012}
\begin{equation}  
 K = \frac{1}{\sum_q \lambda_q^4}
\label{17}
\end{equation}
using eigenvalues $ \lambda_q $ of the Schmidt decomposition of
two-photon amplitude normalized such that $ \sum_q \lambda_q^2 = 1
$.

On the other hand, the number $ K^\Delta_b $ of modes constituting
field $ b $, $ b=s,i $, in a given variable is quantified by the
ratio of appropriate intensity width $ \Delta n_b $ of field $ b $
and width $ \Delta A_b^a $ of the amplitude auto-correlation
function introduced in Eq.~(\ref{15}):
\begin{equation} 
 K^\Delta_b = \frac{\Delta n_b}{\Delta A_b^a} ,
  \hspace{5mm} b=s,i.
\label{18}
\end{equation}

\section{Spatial properties of weak twin beams}

We consider a BBO crystal 8-mm long cut for non-collinear type-I
process (eoo) for the spectrally-degenerate interaction among the
wavelengths $ \lambda_p^0 = 349 $~nm and $ \lambda_s^0 =
\lambda_i^0 = 698 $~nm ($ \vartheta_{\rm BBO} = 36.3 $~deg). The
pump field is provided by the third harmonics of the Nd:YLF laser
at the wavelength 1.047~$\mu $m. Both the Gaussian transverse
profile with radius $ w_p $ and the Gaussian spectrum of the pump
pulse with duration $ \tau_p $ [$ E^\parallel_p(\omega_p) =
\sqrt{\tau_p /\sqrt{2\pi} }
\exp(-\tau_p^2(\omega_p-\omega_p^0)^2/4) $] are considered in the
calculations. Assuming the pump field at normal incidence, the
signal and idler fields at the central frequencies $ \omega_s^0 $
and $ \omega_i^0 $ propagate outside the crystal under the radial
emission angles $ \vartheta_s = \vartheta_i = 8.45 $~deg. As this
configuration is symmetric with respect to the exchange of the
signal and idler fields, we restrict our discussion to only the
signal field. We assume in the discussion that the conditions are
such that the spectral and spatial properties of the twin beams
factorize.

\begin{figure}         
 \centerline{\resizebox{.8\hsize}{!}{\includegraphics{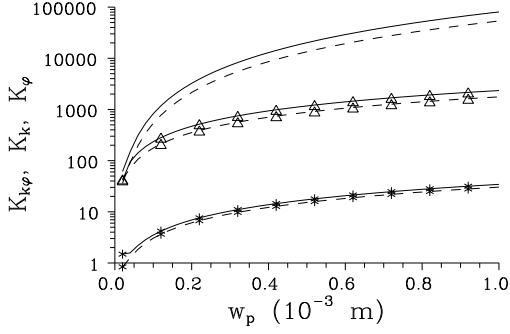}} }
  \caption{Number $ K_{k\varphi} $ of modes in the transverse wave-vector plane
  (plain curve), number $ K_{k} $ of modes in radial direction
  (solid curve with $ \ast $) and number $ K_{\varphi} $ of modes in
  azimuthal direction (solid curve with $ \triangle $) as
  functions of pump-field radius $ w_p $. Dashed curves give the corresponding
  numbers $ K^\Delta_s $ of signal-field modes determined by Eq.~(\ref{18}).}
  \label{fig1}
\end{figure}

\begin{figure}         
 \centerline{\resizebox{.8\hsize}{!}{\includegraphics{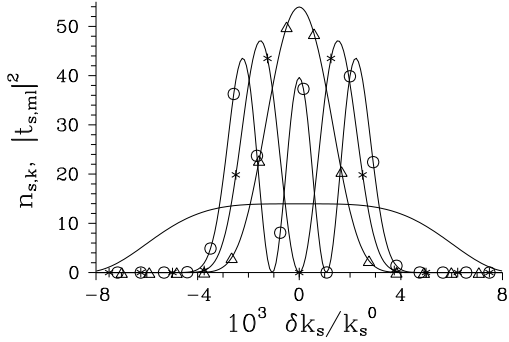}}}
  \caption{Radial intensity profile $ n_{s,k} $ (plain curve) and
  intensity profiles $ |t_{s,0l}|^2 $ of modes for $ l=0 $ (solid curve with
  $ \triangle $), $ l=1 $ (solid curve with $ \ast $) and
  $ l=2 $ (solid curve with $ \circ $) in radial direction of
  the signal field; $ w_p = 1 \times 10^{-3} $~m. Functions are normalized
  such that $ \int dk_s n_{s,k}(k_s)/k_s^0 =1 $ and
  $ \int dk_s |t_{s,0l}(k_s)|^2/k_s^0 =1 $.}
 \label{fig2}
\end{figure}

We first pay attention to the properties of twin beams in the
wave-vector transverse plane. We assume in accord with the
developed model that the twin beam has the rotational symmetry
around the $ z $ axis and so it is stationary in the azimuthal
angle $ \varphi $. Formula (\ref{5}) suggests that the rotational
symmetry is roughly observed for the pump-field radius $ w_p $
larger than $ w_p^a $ that equals 270~$ \mu $m for the 8-mm long
crystal and the considered geometry [$ n_p = 1.658 $, $
|dn_p/d\vartheta_p| = 0.123 $]. The pump-field radius $ w_p $ is
the crucial parameter that determines the properties of twin beams
in the transverse plane. It gives the number $ K_{k\varphi} $ of
independent transverse modes comprising the twin beam. In our
configuration, the greater the value of radius $ w_p $ the greater
the number $ K_{k\varphi} $ of independent modes, as documented in
Fig.~\ref{fig1}.
The number $ K_{k\varphi} $ of overall modes in the transverse
wave-vector plane can approximately be factorized into the number
$ K_k $ of modes in the radial direction and the number $
K_\varphi $ of modes in the azimuthal direction. This
factorization arises from the fact that the number $ K $ of radial
modes determined for a fixed value of the azimuthal number $ m $
depends only weakly on the number $ m $. The number $ K_\varphi $
of azimuthal modes is much larger than the number $ K_k $ of
radial modes as these modes cover the whole circle. There
typically occur hundreds or thousands of independent azimuthal
modes. On the other hand, from several up to several tens of
radial modes are needed for the description of a typical twin
beam. Whereas the azimuthal modes take the form of harmonic
functions, damped exponential functions modified by oscillating
polynomials \cite{PerinaJr2008} approximate the radial modes. The
first couple of radial modes for $ m=0 $ are shown in
Fig.~\ref{fig2} together with the radial intensity profile $
n_{s,k} $ for the pump field 1-mm wide.
It holds that the intensity profile of an $ l $-th radial mode has
$ l $ zeroes and $ l+1 $ local peaks. Also the larger the value of
$ l $, the wider the mode. However and importantly, all modes are
distributed along the whole intensity profile $ n_{s,k}(k_s) $. We
note that, in Fig.~\ref{fig1}, for the values of pump-field radius
$ w_p $ smaller than 0.270~$ \mu $m the generated twin beams are
influenced by the crystal anisotropy. Anisotropy leads in general
to more strict phase-matching conditions which results in greater
values of the Schmidt numbers $ K $ compared to those shown in
Fig.~\ref{fig1}. Anisotropy and its role in photon-pair generation
has been discussed in detail in \cite{Fedorov2008}.

Numbers $ K^\Delta_s $ of modes in the signal field determined
along Eq.~(\ref{18}) are compared with the numbers $ K $ of paired
modes from the Schmidt decomposition in Fig.~\ref{fig1}. This
comparison reveals that the values of $ K^\Delta $ are only
approx. by 40\% lower than the values of $ K $. The numbers $
K^\Delta $ of effectively populated signal-field modes are in
general smaller than the corresponding Schmidt numbers $ K $ as
the widths of auto-correlation functions occurring in definition
(\ref{18}) are not able to fully take into account the complex
structure of entanglement present in a twin beam
\cite{Chan2007,Just2013}. In other words, numbers $ K $ of paired
modes and numbers $ K^\Delta_s $ of signal-field modes are defined
differently. However, their comparison done in Fig.~\ref{fig1}
clearly reveals that both of them are suitable for quantifying the
dimensionality of a twin beam.

\begin{figure}         
 \centerline{\resizebox{0.47\hsize}{!}{\includegraphics{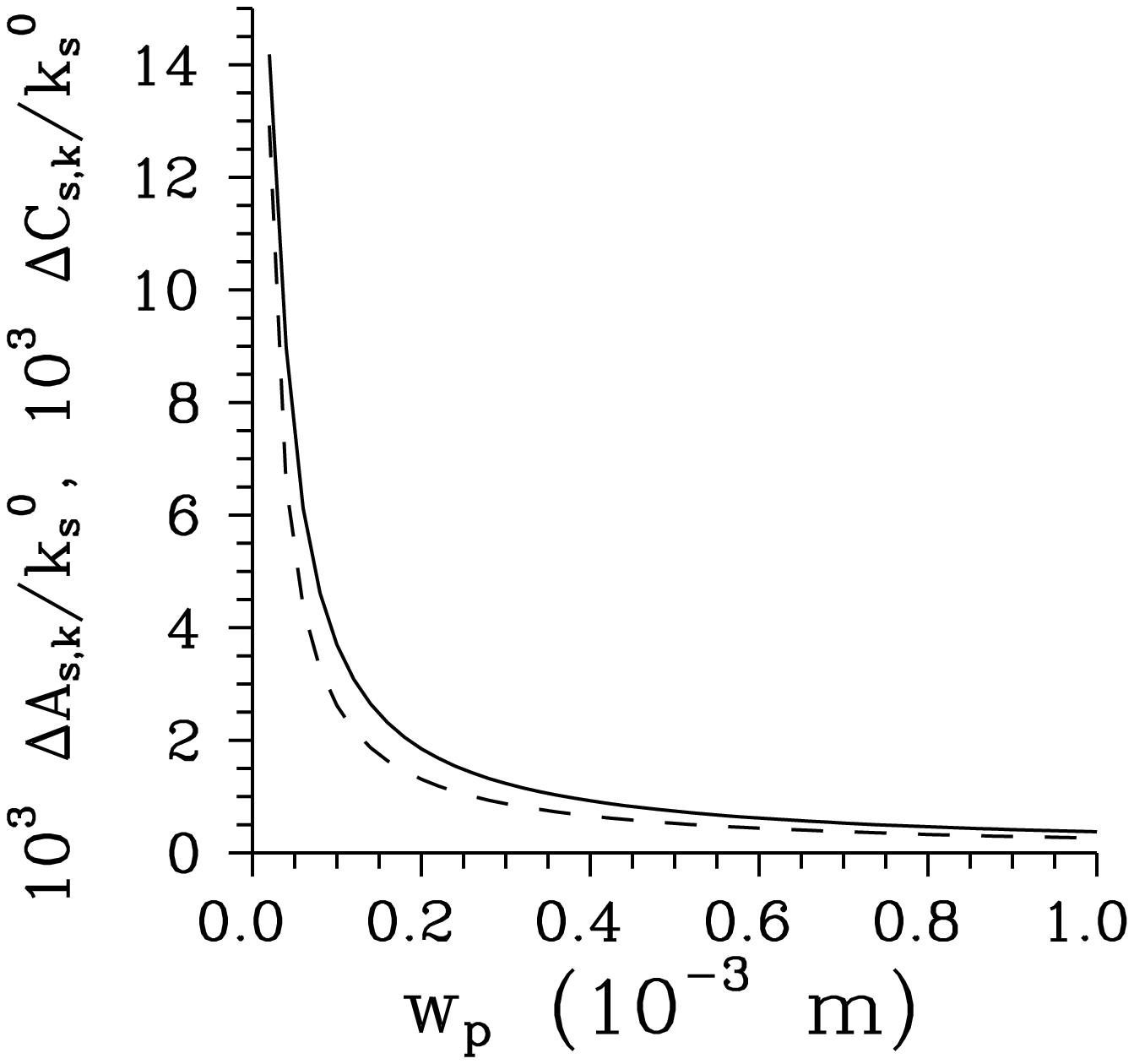}}
  \hspace{1mm}
 \resizebox{0.47\hsize}{!}{\includegraphics{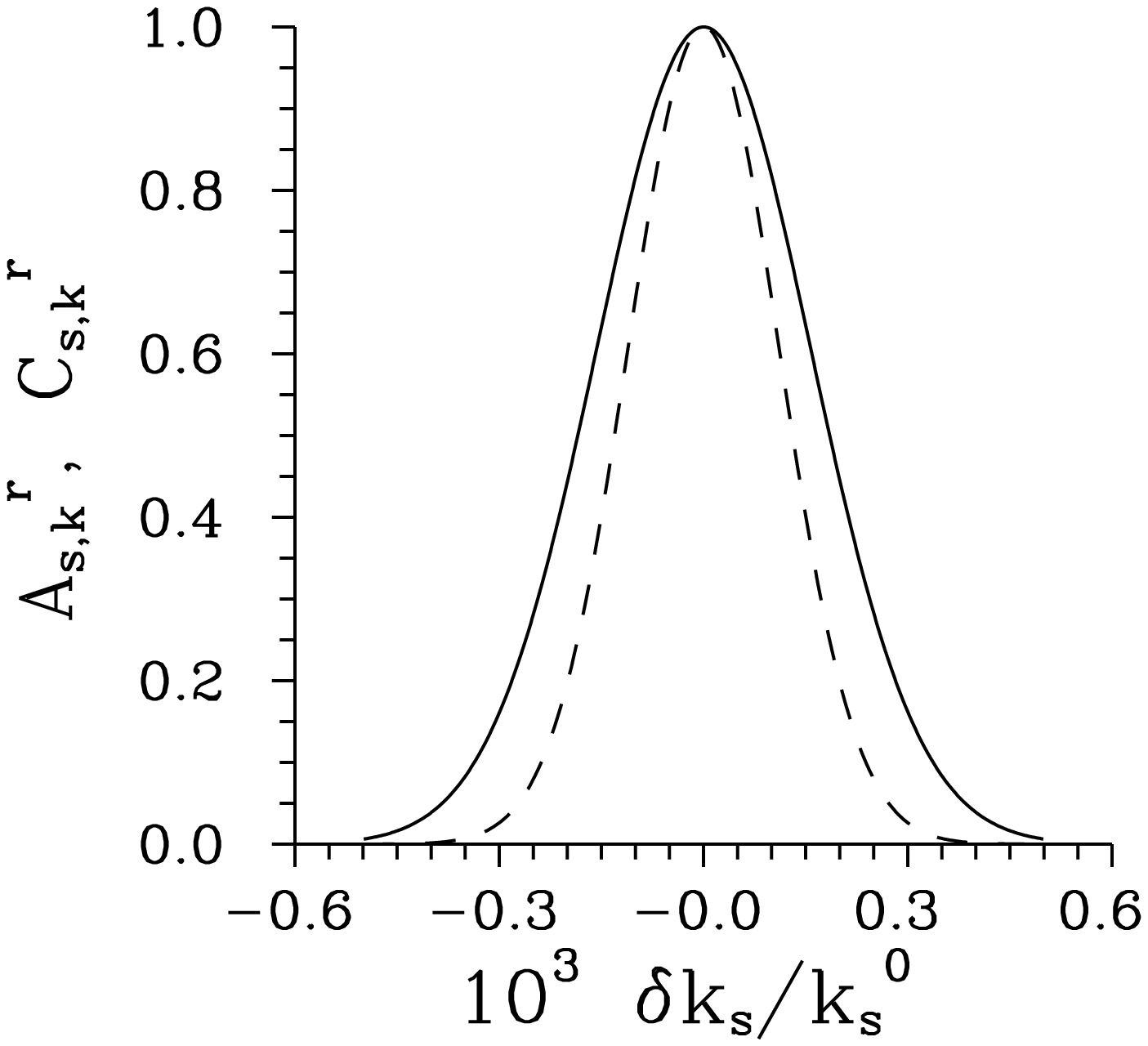}}}

 \centerline{(a) \hspace{.5\hsize} (b)}

 \caption{(a) Widths $ \Delta A_{s,k} $ (FWHM, full width at half maximum)
  of signal-field intensity auto-correlation
  function (plain curve) and $ \Delta C_{s,k} $ of intensity cross-correlation function
  (dashed curve) in radial direction as they depend on pump-field radius $ w_p $.
  In (b), functions $ A_{s,k}^r(\delta k_s) \equiv A_{s,k}(k_s^0+\delta k_s,k_s^0)/
  A_{s,k}(k_s^0,k_s^0) $ and $ C_{s,k}^r(\delta k_s) \equiv C_{s,k}(k_s^0+\delta k_s,k_i^0)/
  C_{s,k}(k_s^0,k_i^0) $ are plotted for $ w_p = 1 \times 10^{-3} $~m.}
 \label{fig3}
\end{figure}

\begin{figure}         
 \centerline{\resizebox{0.47\hsize}{!}{\includegraphics{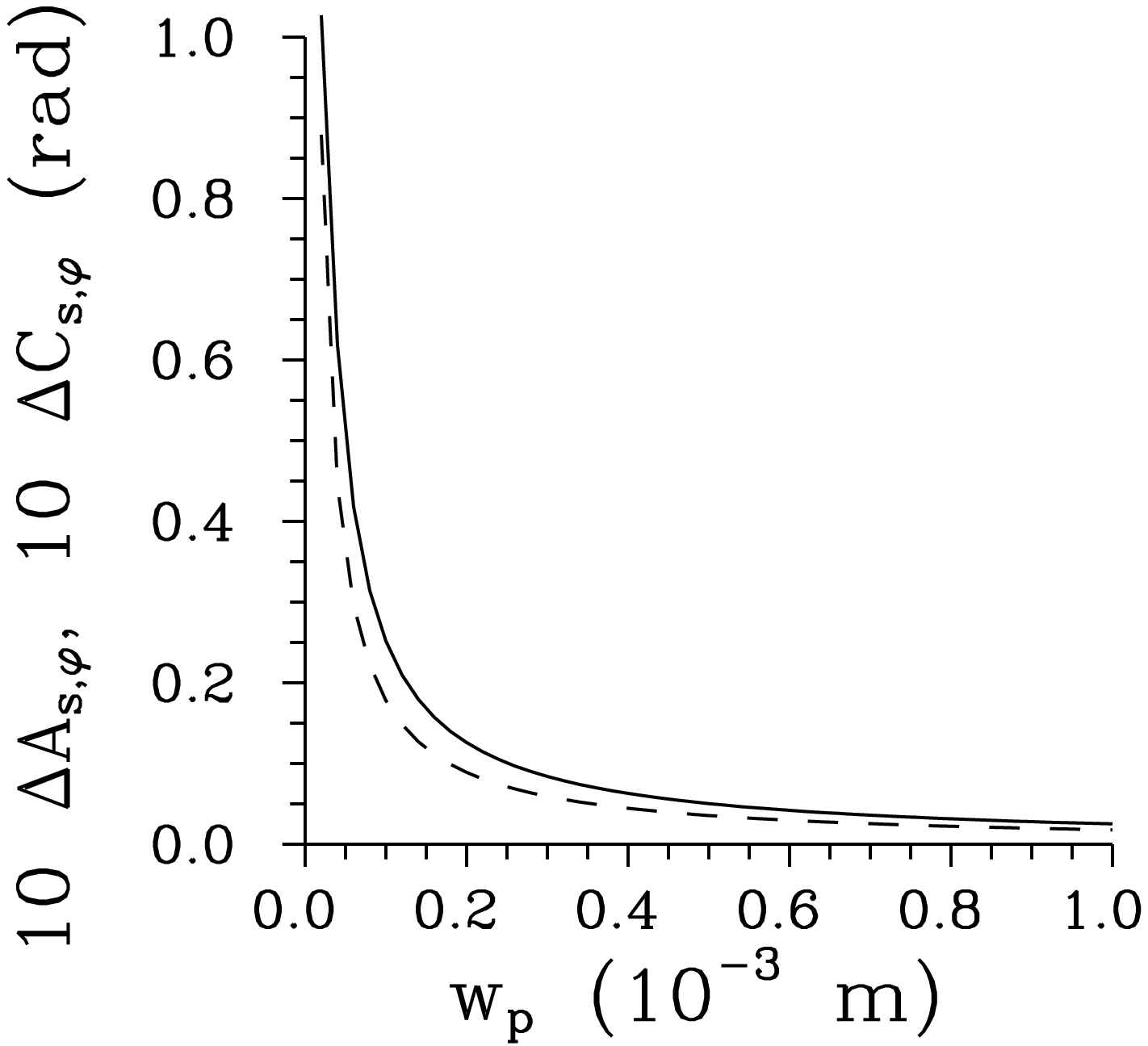}}
  \hspace{1mm}
 \resizebox{0.47\hsize}{!}{\includegraphics{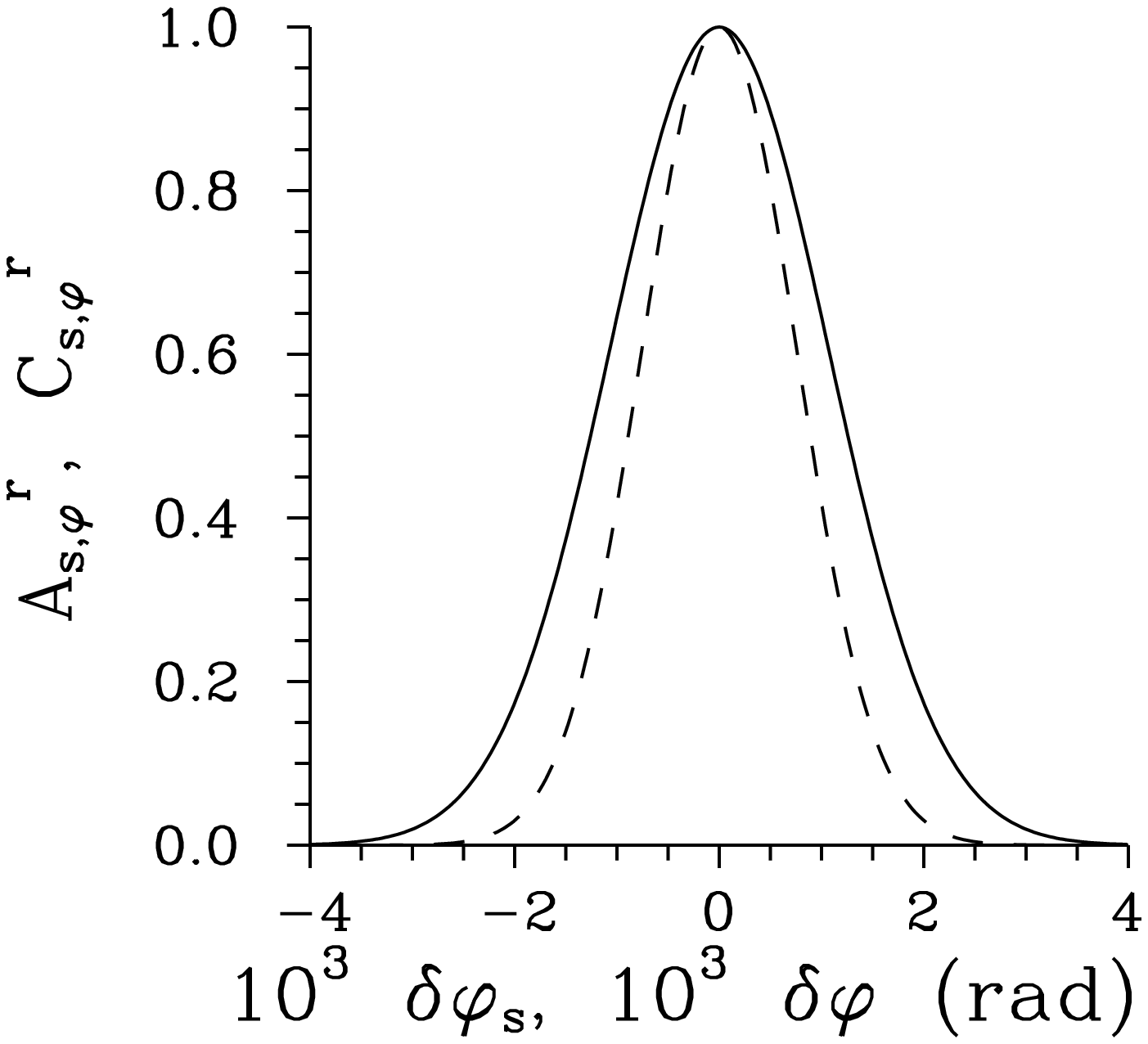}}}

 \centerline{(a) \hspace{.5\hsize} (b)}

 \caption{(a) Widths $ \Delta A_{s,\varphi} $ (FWHM) of signal-field intensity auto-correlation
  function (plain curve) and $ \Delta C_{s,\varphi} $ of intensity cross-correlation function
  (dashed curve) in azimuthal direction depending on pump-field radius $ w_p $.
  In (b), intensity auto-correlation function $ A_{s,\varphi}^r(\delta\varphi_s) \equiv
  A_{s,\varphi}(\varphi_s^0+\delta\varphi_s,\varphi_s^0) / A_{s,\varphi}(\varphi_s^0,\varphi_s^0) $
  valid for an arbitrary angle $ \varphi_s^0 $ and cross-correlation function $ C_{s,\varphi}^r(\delta\varphi)
  \equiv C_{s,\varphi}(\varphi_s^0+\delta\varphi,\varphi_i^0) /
  C_{s,\varphi}(\varphi_s^0,\varphi_i^0) $ given for arbitrary angles fulfilling $ \varphi_i^0 =\varphi_s^0+\pi $
  are shown for $ w_p = 1 \times 10^{-3} $~m.}
 \label{fig4}
\end{figure}

As the range of radial angles $ \vartheta_s $ belonging to the
emitted photons practically does not change with the pump-field
radius $ w_p $ and the allowed azimuthal emission angles $
\varphi_s $ lie in interval $ (0,2\pi) $, an increase of the
number $ K^\Delta_{s,k\varphi} $ of signal-field modes with the
radius $ w_p $ is caused by the decrease of the radial and
azimuthal widths of effective modes given by the extension of
amplitude auto-correlation functions $ A_{s,k}^a $ and $
A_{s,\varphi}^a $ defined in Eqs.~(\ref{15}). The dependence of
widths $ \Delta A_{s,k} $ and $ \Delta A_{s,\varphi} $ of their
intensity counterparts on the radius $ w_p $ is shown in
Figs.~\ref{fig3}(a) and \ref{fig4}(a), respectively, and confirms
this behavior.
Profiles $ A_{s,k} $ and $ A_{s,\varphi} $ giving the signal-field
intensity correlations for the pump field 1-mm wide are drawn in
Figs.~\ref{fig3}(b) and \ref{fig4}(b).

The generation of signal and idler fields by photon pairs strongly
correlated in the transverse plane results in strong
cross-correlations between the intensities of the signal and idler
fields. Also these cross-correlations described by the
cross-correlation functions $ C_{s,k} $ and $ C_{s,\varphi} $
given in Eqs.~(\ref{16}) depend strongly on the pump-field radius
$ w_p $. The larger the radius $ w_p $ the smaller the widths $
\Delta C_{s,k} $ and $ \Delta C_{s,\varphi} $ of intensity
cross-correlation functions in the radial and azimuthal
directions, respectively [see Figs.~\ref{fig3}(a) and
\ref{fig4}(a)] \cite{Hamar2010,Machulka2014}. The comparison of
widths $ \Delta C_{s,k} $ and $ \Delta A_{s,k} $ (or $ \Delta
C_{s,\varphi} $ and $ \Delta A_{s,\varphi} $) plotted in
Figs.~\ref{fig3}(a) and \ref{fig4}(a) reveals that the intensity
cross-correlation functions are narrower than the intensity
auto-correlation functions. This behavior can qualitatively be
explained as follows. Cross-correlations are formed by individual
photon pairs with certain correlations between the signal and
idler photons. As these correlations reflect the geometry of the
nonlinear process, blurring of the ideal point-like correlations
(in the transverse wave-vector space) occurs. On the other hand,
intensity auto-correlations in the signal (or idler) field are
created at the quantum level by at least two photon pairs. Roughly
speaking, the first emitted photon pair provides a signal photon
and its correlated idler twin. The annihilation of the second
photon pair using the already present idler photon then creates
photon-photon auto-correlations in the signal field. As we need
two photon pairs in this reasoning, the overall blurring is larger
compared to that of one photon pair. As a consequence, the
intensity auto-correlation functions are wider than the intensity
cross-correlation functions. Also similarity of the profiles of
intensity auto-correlation functions $ A_{s,k} $ and $
A_{s,\varphi} $ and intensity cross-correlation functions $
C_{s,k} $ and $ C_{s,\varphi} $, respectively, can be deduced from
this reasoning. The curves plotted in Figs.~\ref{fig3}(b) and
\ref{fig4}(b) confirm this behavior.

\section{Spectral properties of weak twin beams}

\begin{figure}         
 \centerline{\resizebox{.8\hsize}{!}{\includegraphics{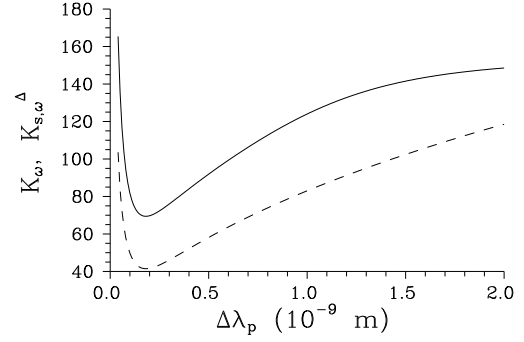}}}
  \caption{Number $ K_{\omega} $ of spectral modes depending on pump-field
  spectral width $ \Delta\lambda_p $; $ \Delta \lambda_p = 4\pi \sqrt{2\ln(2)} c /
  [(\omega_p^{0})^2\tau_p] $. Dashed curve gives the
   number $ K^\Delta_{s,\omega} $ of modes according to Eq.~(\ref{18}).}
  \label{fig5}
\end{figure}

The behavior of twin beam in the spectral domain and under the
considered conditions is more complex compared to the spatial
domain. The number $ K_\omega $ of paired spectral modes attains
its minimum when considered as a function of the pump-field
spectral width $ \Delta \lambda_p $ \cite{Mikhailova2008}. The
number $ K_\omega $ of paired spectral modes is larger ($ K_\omega
\approx 70 $) even in this minimum reached for the pump pulse with
the spectrum approx. 0.2~nm wide, as documented in
Fig.~\ref{fig5}.
It holds also in the spectral domain that the number $
K_\omega^\Delta $ of signal-field modes determined from the ratio
defined in Eq.~(\ref{18}) is smaller than the number $ K_\omega $
of paired modes arising from the Schmidt decomposition (see
Fig.~\ref{fig5}). The comparison of curves in Fig.~\ref{fig5}
confirms that both the number $ K_\omega $ of paired modes and the
number $ K_\omega^\Delta $ of signal-field modes are suitable for
quantifying dimensionality of the twin beam in a broad range of
pump-field spectral widths.

\begin{figure}         
 \centerline{\resizebox{0.47\hsize}{!}{\includegraphics{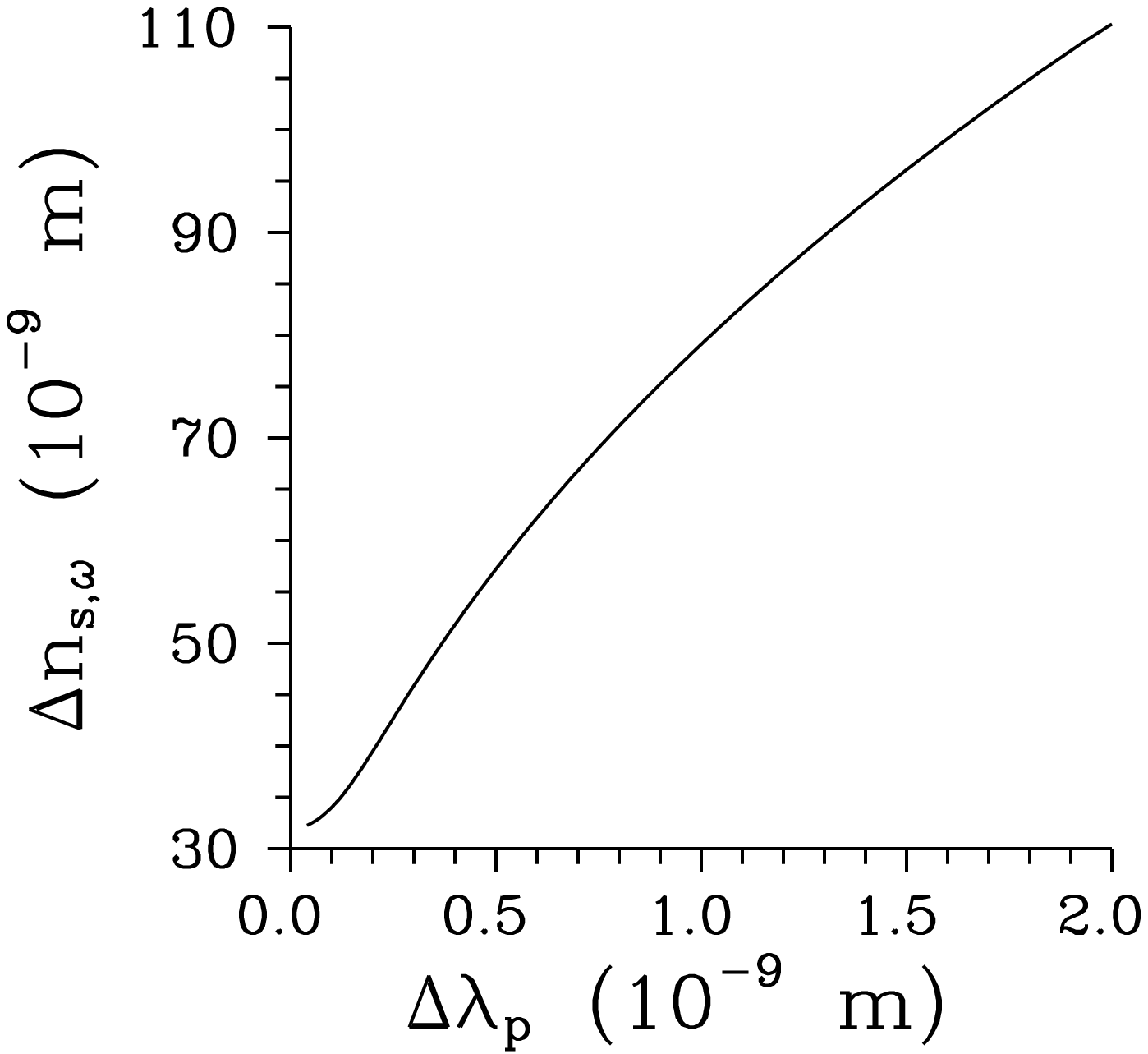}}
  \hspace{1mm}
 \resizebox{0.47\hsize}{!}{\includegraphics{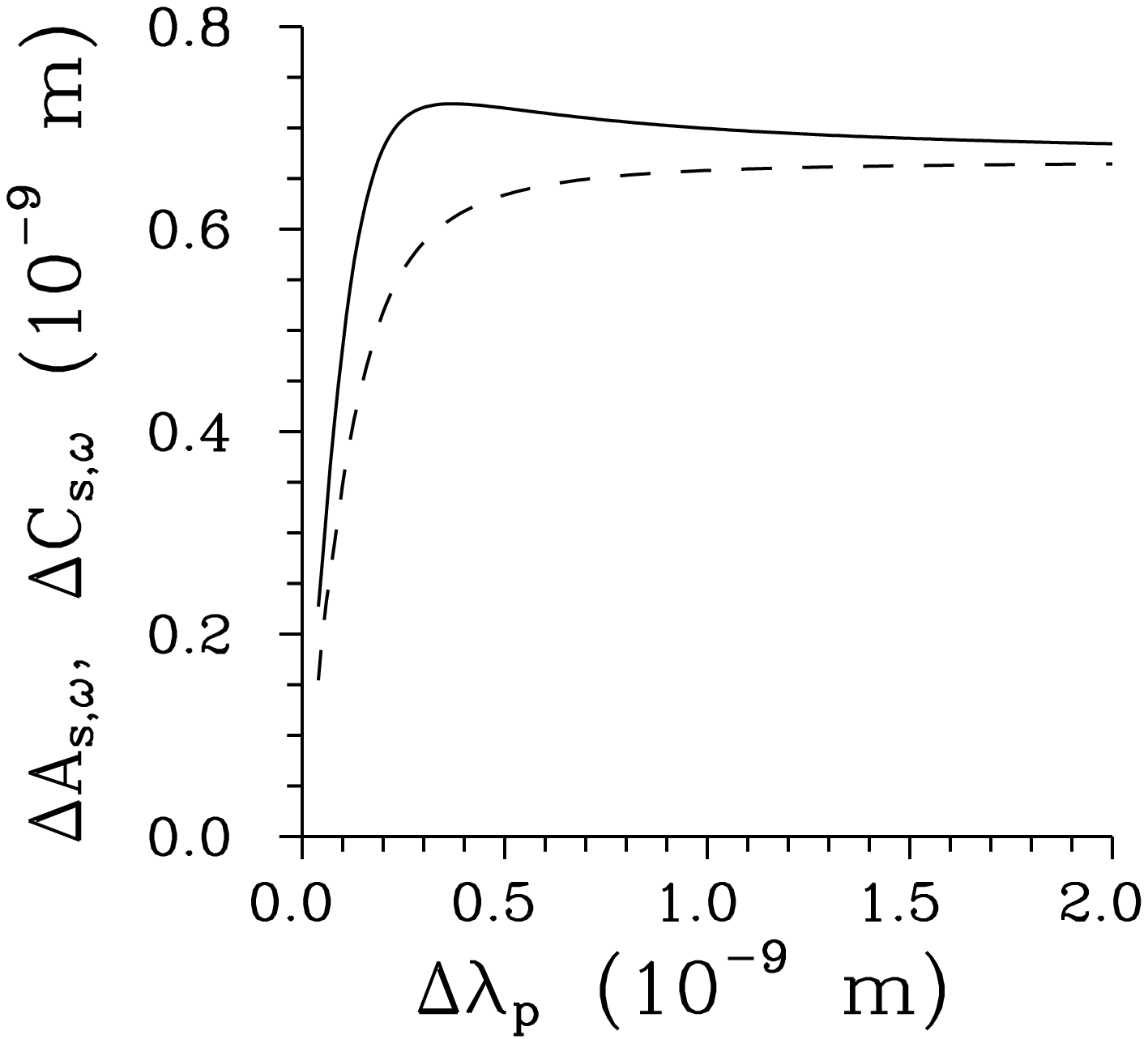}}}

 \centerline{(a) \hspace{.5\hsize} (b)}

 \caption{(a) Width $ \Delta n_{s,\omega} $ (FWHM) of signal-field intensity
  spectrum and (b) widths $ \Delta A_{s,\omega} $ (FWHM) of signal-field
  intensity auto-correlation function (plain curve) and $ \Delta C_{s,\omega} $ of intensity
  cross-correlation function (dashed curve) as they depend on pump-field
  spectral width $ \Delta\lambda_p $.}
\label{fig6}
\end{figure}

\begin{figure}         
 \centerline{\resizebox{0.47\hsize}{!}{\includegraphics{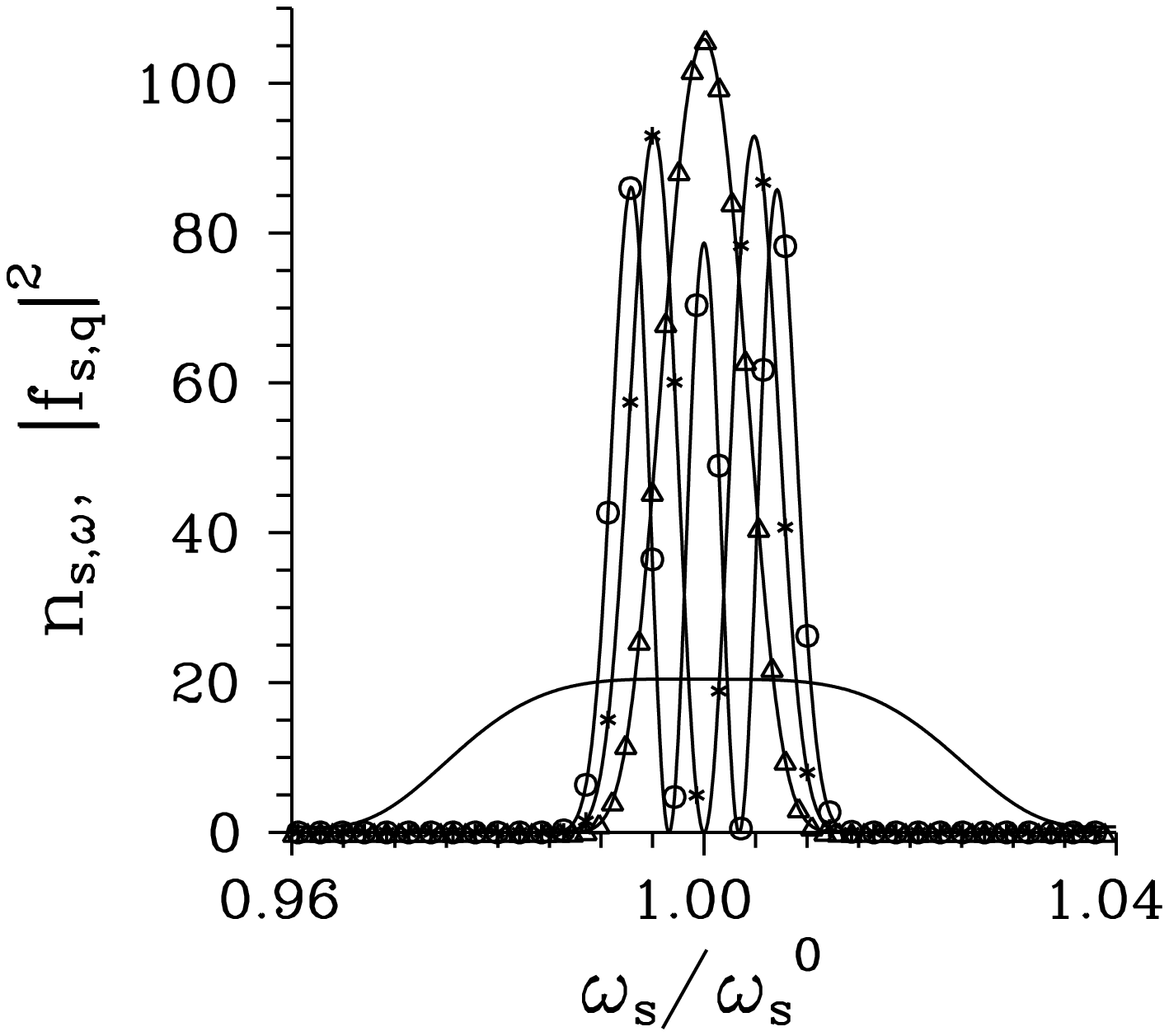}}
  \hspace{1mm}
 \resizebox{0.47\hsize}{!}{\includegraphics{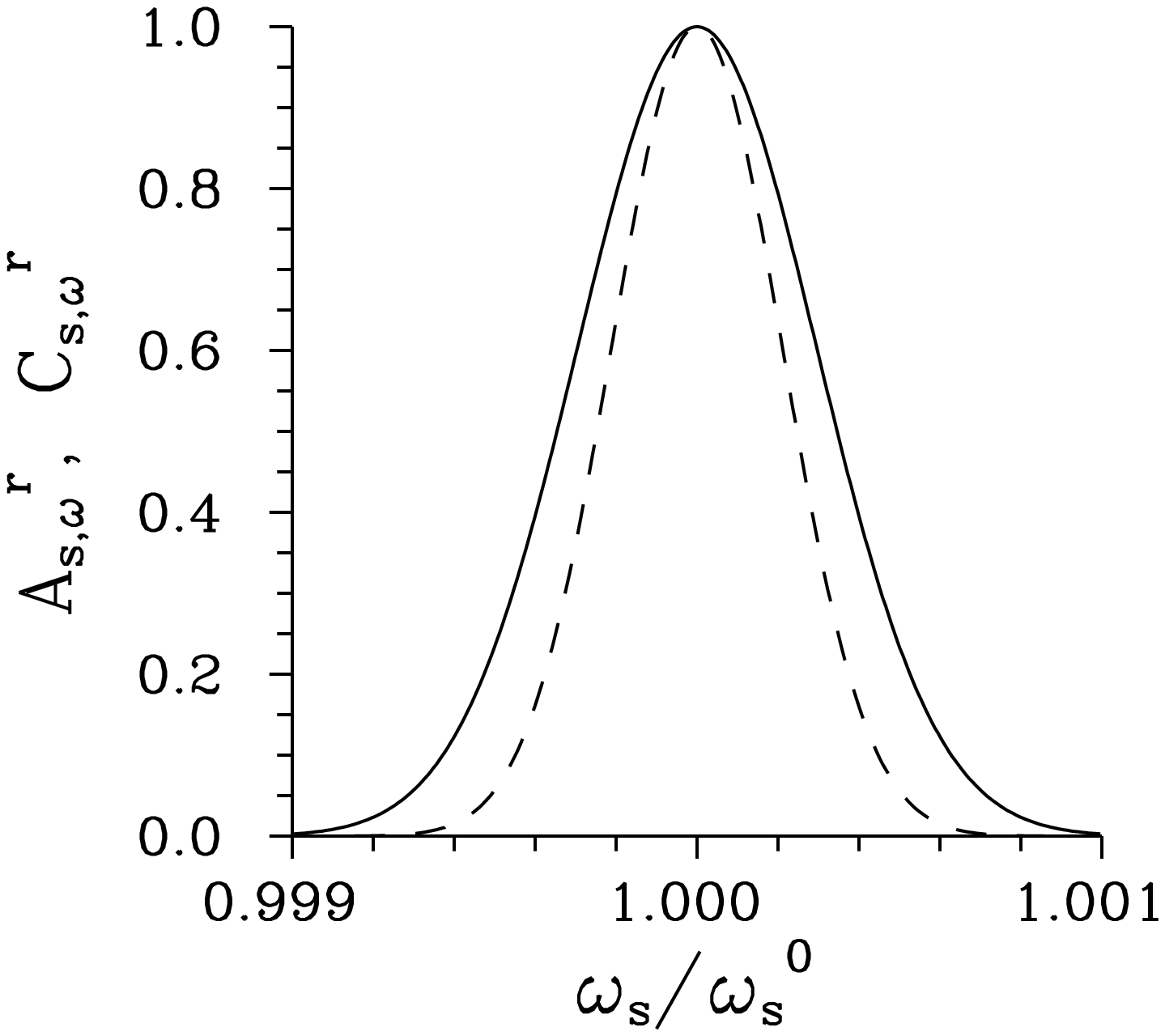}}}

 \centerline{(a) \hspace{.5\hsize} (b)}

 \caption{(a) Spectral intensity $ n_{s,\omega} $ (plain curve) and
  spectral intensities $ |f_{s,q}|^2 $ of modes for $ q=0 $ (solid curve with
  $ \ast $), $ q=1 $ (solid curve with $ \triangle $) and
  $ q=2 $ (solid curve with $ \circ $) in the signal field. (b) Signal-field
  spectral intensity auto-correlation
  function $ A_{s,\omega}^r(\omega_s) \equiv A_{s,\omega}(\omega_s,\omega_s^0)/
  A_{s,\omega}(\omega_s^0,\omega_s^0) $ (plain curve) and cross-correlation
  function $ C_{s,\omega}^r(\omega_s) \equiv C_{s,\omega}(\omega_s,\omega_i^0)/
  C_{s,\omega}(\omega_s^0,\omega_i^0) $ (dashed curve); $ \Delta\lambda_p
  = 1 \times 10^{-10} $~m. Normalization is such that $ \int d\omega_s
  n_{s,\omega}(\omega_s)/\omega_s^0 = 1 $ and $ \int d\omega_s
  |f_{s,q}(\omega_s)|^2/\omega_s^0 = 1 $.}
 \label{fig7}
\end{figure}

The dependence of the number $ K_\omega^\Delta $ of signal-field
modes on the pump-field spectral width $ \Delta\lambda_p $ as
drawn in Fig.~\ref{fig5} can be explained by the behavior of
spectral intensity width $ \Delta n_{s,\omega} $ and width $
\Delta A_{s,\omega} $ of intensity auto-correlation function.
Whereas the spectral intensity width $ \Delta n_{s,\omega} $
monotonically increases as a function of $ \Delta\lambda_p $ [see
Fig.~\ref{fig6}(a)], the width $ \Delta A_{s,\omega} $ of
intensity auto-correlation function increases for lower values of
$ \Delta\lambda_p $ and then it saturates [see
Fig.~\ref{fig6}(b)]. This results in an increase of the values of
$ K_\omega^\Delta $ for larger values of $ \Delta\lambda_p $
observed in Fig.~\ref{fig5}.
Modes of the spectral decomposition behave in the same manner as
the modes in the radial direction. Thus, the intensity profile of
a $ q $-th mode has $ q $ zeros and $ q+1 $ peaks and extends over
all frequencies found in the spectrum [see Fig.~\ref{fig7}(a)].
As a consequence of the conservation law of energy, the
frequencies of the emitted signal and idler photons are correlated
within the spectral interval given by the pump-field spectral
width. However, the interval of correlations is also influenced by
the phase-matching conditions that limit its extension. It follows
that a certain width of this interval is reached with the
increasing pump-field spectral width $ \Delta\lambda_p $ and no
further increase is possible [see Fig.~\ref{fig6}(b)]. This
determines correlations between the intensities of the signal and
idler fields as well as among the intensities inside the
individual fields. The same argumentation as that used in the
transverse wave-vector domain shows that the intensity
auto-correlation function $ A_{s,\omega} $ has to be broader than
the intensity cross-correlation function $ C_{s,\omega} $ [compare
the curves in Fig.~\ref{fig6}(b)] and also that profiles of these
correlation functions plotted in Fig.~\ref{fig7}(b) are similar.

The picture that the overall spectrum is composed of adjacent
independent 'local' modes is useful in predicting the behavior of
numbers $ K_\omega $ of paired modes and $ K^\Delta_{s,\omega} $
of signal-field modes with respect to spectral filtering. It
suggests that the narrower the filter width is, the lower the
numbers $ K_\omega $ and $ K^\Delta_{s,\omega} $ of modes needed
in the description of a twin beam. This behavior has been
confirmed numerically. A sufficiently strong filtering then allows
to reach a twin beam with the number of modes approaching one
\cite{Perez2014}. The same conclusions are valid for the geometric
filtering that reduces the numbers $ K_{k\varphi} $ and $
K_{s,k\varphi}^\Delta $ of modes in the transverse wave-vector
plane.

\section{Conclusions}

Weak spatio-spectral twin beams have been analyzed in paraxial
approximation using the perturbation solution of the
Schr\"{o}dinger equation and its decomposition into the spatial
and spectral paired modes. Applying these modes, coherence
properties of weak twin beams have been studied using both the
auto- and cross-correlation functions. Numbers of paired modes
revealed by the Schmidt decomposition and numbers of modes
constituting the signal (or idler) field and given by the ratio of
field width and width of the appropriate amplitude
auto-correlation function have been found mutually proportional in
the broad area of the analyzed pump-field parameters. This
justifies both of them as appropriate quantifiers of
dimensionality of weak twin beams. Whereas the number of
transverse modes increases with the increasing pump-field radius
in our configuration, the number of spectral modes considered as a
function of pump-field spectral width has a well-formed minimum.
This behavior has been explained analyzing the widths of
appropriate correlation functions. It has been shown that the
spectral and transverse wave-vector cross-correlation functions
are broader than the corresponding auto-correlation functions.

\ack

The author thanks M. Bondani, O. Haderka and A. Allevi for
stimulating discussions. He gratefully acknowledges the support by
project LO1305 of the Ministry of Education, Youth and Sports of
the Czech Republic.

\section*{References}


\end{document}